\documentclass[12pt]{article}
\usepackage{layout,amssymb,amsmath,epsfig}
\allowdisplaybreaks
\begin{document}

\title{Cosmological Implications of the Generalized Entropy Based Holographic Dark Energy Models in Dynamical Chern-Simons Modified Gravity}
\author{\textbf{M. Younas}$^{1}$ \thanks {muhammadyounas@cuilahore.edu.pk},
\textbf{Abdul Jawad}$^{1}$\thanks {abduljawad@cuilahore.edu.pk},
\textbf{Saba Qummer}$^1$\thanks {sabaqummer143@gmail.com},\\
\textbf{H. Moradpour}$^2$\thanks{h.moradpour@riaam.ac.ir} and
\textbf{Shamaila Rani} $^1$ \thanks {drshamailarani@cuilahore.edu.pk}\\
$^1$Department of Mathematics, COMSATS University Islamabad,\\
Lahore Campus-54000, Pakistan.\\ $^2$Research Institute for
Astronomy and Astrophysics\\ of Maragha (RIAAM), P.O. Box 55134-441,
Maragha, Iran.}

\maketitle
\begin{abstract}
Recently, Tsallis, R\'{e}nyi and Sharma-Mitall and entropies have
widely been used to study the gravitational and cosmological setups.
We consider a flat FRW universe with linear interaction between dark
energy and dark matter. We discuss the dark energy models using
Tsallis, R\'{e}nyi and Sharma-Mitall entropies in the framework of
Chern-Simons modified gravity. We explore various cosmological
parameters (equation of state parameter, squared sound of speed )
and cosmological plane ($\omega_{d}-\omega_{d}'$, where
$\omega_{d}'$, is the evolutionary equation of state parameter). It
is observed that the equation of state parameter gives
quintessence-like nature of the universe in most of the cases. Also,
the squared speed of sound shows stability of the models for
Tsallis, R\'{e}nyi dark energy model while unstable behavior for
Sharma-Mitall dark energy model. The $\omega_{d}-\omega_{d}'$ plane
represents the thawing region for all dark energy models.
\end{abstract}

\section{Introduction}

In last few years, a remarkable progress have seen in understanding
of the universe expansion. It has been approved by current
observational data that the universe undergoes an accelerated
expansion. The observations of type Ia Super Novae
(SNeIa)\cite{1}-\cite{4}, large scale structure (LSS)
\cite{5}-\cite{8} and Cosmic Microwave Background Radiation (CMBR)
\cite{9}-\cite{11}, determined that the expansion of the universe is
currently accelerating. It is also consensus that this  acceleration
is generally believed to be caused by a mysterious form of energy or
exotic matter with negative pressure so called dark energy (DE)
\cite{12}-\cite{23}.

The discovery of accelerating expansion of the universe is a
milestone for cosmology. It is considered that $95\%$ of our
universe is composed of two components, that is DE and dark matter
\cite{17}. The dark matter constitutes about $25\%$ of the total
energy density of the universe. The existence of the universe is
proved by astrophysical observation but the nature of dark matter is
still unknown. Mainly the DE is also a curious component of our
universe. It is responsible for current accelerating universe and DE
is entirely different from baryonic matter. DE constitutes almost
$70\%$ of the total energy density of our universe.

In order to describe the accelerated expansion phenomenon, two
different approaches have been adopted. One is the proposal of
various dynamical DE models such as family of Chaplygin gas,
holographic, Quintessence, K-essence, Ghost etc \cite{17}. A second
approach for understanding this strange component of the universe is
modifying the standard theories of gravity, namely, general
relativity (GR). Several modified theories of gravity are $f(R),
f(T) \cite{18}, f(R, \textrm{T} ) \cite{19}$,  $f(G) \cite{20}$,
where $R$ is the curvature scalar, $T$ denotes the torsion scalar,
$\textrm{T}$ is the trace of the energy momentum tensor and $G$ is
the invariant of Gauss-Bonnet.

Holographic DE (HDE) model is favorable technique to solve DE
mystery which has arisen a lot of attentions and is based upon the
holographic principle that states the number of degrees of freedom
of a system scales with its area instead of its volume. In fact, HDE
relates the energy density of quantum fields in vacuum (as the DE
candidate) to the infrared and ultraviolet cutoffs. In addition, HDE
is an interesting effort in exploring the nature of DE in the
framework of quantum gravity. Cohen et al. \cite{33}, studied that
the construction of HDE density is based on the relation about the
vacuum energy of the system whose maximum amount should not exceed
the black hole mass. Cosmological consequences of some HDE models in
the dynamical Chern-Simons framework, as a modified gravity theory,
can be found in Ref \cite{34}.

By considering the long term gravity with the nature of spacetime,
different entropy formalism have been used to observe the
gravitational and cosmological effects \cite{29,28,30,24}. The HDE
models such as Tsallis HDE (THDE) \cite{28}, R\'{e}nyi HDE model
(RHDE) \cite{30} and Sharma-Mitall HDE (SMHDE) \cite{24} have been
recently proposed. In the standard cosmology framework, and from the
classical stability view of point, while THDE is not stable
\cite{28}, RHDE is stable during the cosmic evolution \cite{30} and
SMHDE is stable only whenever it becomes dominant in the world
\cite{24}. In the present work, we use the Tasllis, Sharam-Mitall
and R\'{e}nyi entropies in the frame work of dynamical Chern-Simons
modified gravity and consider an interaction term. We investigate
the different cosmological parameters such as equation of state
parameter, the cosmological $\omega_{d}-\omega_{d}'$ plane where,
$\omega_{d}'$ shows the evaluation with respect to $\ln a$. We also
investigate the squared sound speed of the HDE model to check the
stability and there graphical approach.

This paper is organized as follows. In section 2, we provide the
basics of Chern-Simons modified gravity. In section 3, we observe
the equation of state parameter (EoS), cosmological plane and
squared sound speed for THDE model. Sections 4 and 5 are devoted to
find the cosmological parameter, cosmological plane and squared of
sound speed for RHDE and SMHDE  models respectively. In the last
section, we conclude the results.

\section{Dynamical Chern-Simons Modified Gravity}

In this section, we give a review of dynamical Chern-Simons modified
gravity. The action which describes the Chern-Simons modified
gravity is given as
\begin{equation}\label{1}
S=\frac{1}{16 \pi G}\int_{\nu}d^{4}x\left[\sqrt{-g}R+\frac{l}{4} \theta {^\ast R}^{\rho\sigma\mu\nu}R_{\rho\sigma\mu\nu}
-\frac{1}{2}g^{\mu\nu}\nabla_{\mu}\theta\nabla_{\nu}\theta +V(\theta)\right]+S_{mat},
\end{equation}
where $R$ represents the Ricci scalar, ${^\ast R}^{\rho\sigma\mu\nu}R_{\rho\sigma\mu\nu}$
is a topological invariant called the Pontryagin term, $l$ is a coupling constant,
$\theta$ shows the dynamical variable, $S_{mat}$ represents the action of matter
and $V(\theta)$ is the potential term. In the case of string theory, we use $V(\theta)=0$.
By varying the action equation with respect to $g_{\mu\nu}$ and the scalar field $\theta$,
we get the following field equations
\begin{eqnarray}
G_{\mu\nu}+l C_{\mu\nu}&=&8 \pi G T_{\mu\nu},\nonumber \\
\label{2}g^{\mu\nu}\nabla_{\mu}\nabla_{\nu}\theta&=&-\frac{l}{64 \pi}{^\ast R}^{\rho\sigma\mu\nu}R_{\rho\sigma\mu\nu}.
\end{eqnarray}
Here, $G_{\mu\nu}$ and $C_{\mu\nu}$ are Einstein tensor and Cotton tensor, respectively. The Cotton tensor  $C_{\mu\nu}$
is defined as
\begin{equation}\label{3}
C_{\mu\nu}=-\frac{1}{2\sqrt{-g}}((\nabla_{\rho}\theta)\varepsilon^{\rho \beta \tau (\mu} \nabla_{\tau} R^{\nu)}_{\beta})
+(\nabla_{\sigma}\nabla_{\rho}\theta)^{\ast} {R}^{\rho(\mu\nu)\sigma}.
\end{equation}
The energy-momentum tensor are given by
\begin{eqnarray}
{\hat{T}^{\theta}}_{\mu\nu}&=&\nabla_{\mu}\theta \nabla_{\nu}\theta-\frac{1}{2}g_{\mu\nu}\nabla^
{\rho}\theta\nabla_{\rho}\theta,\nonumber \\
\label{4}T_{\mu\nu}&=&(\rho+p)u_{\mu}u_{\nu}+pg_{\mu\nu},
\end{eqnarray}
where, $T_{\mu\nu}$ shows the matter contribution and ${\hat{T}^{\theta}}_{\mu\nu}$
represents the scalar field contribution while, $P$ and $\rho$ represent the pressure and
energy density respectively. Furthermore, $u_{\mu}=(1,0, 0, 0)$ is the four velocity.
In the frame work of Chern-Simons gravity, we get
the following Friedmann equation
\begin{equation}\label{5}
H^{2}=\frac{1}{3}(\rho_{m}+\rho_{d})+\frac{1}{6}\dot{\theta}^{2},
\end{equation}
where, $H=\frac{\dot{a}}{a}$ is the Hubble parameter and dot represents the derivative of
$a$ with respect to $t$ and $8\pi G=1$.
For FRW spacetime, the ponytrying term $^{^\ast}RR$ vanishes identically
therefore, the scalar field in Eq.(\ref{2}) takes the following form
\begin{equation}\label{6}
g^{\mu\nu}\nabla_{\mu}\nabla_{\nu}\theta=g^{\mu\nu}[\partial_{\nu}\partial_{\mu}\theta]=0.
\end{equation}
We set, $\theta=\theta(t)$ and get the following equation
\begin{equation}\label{7}
\ddot{\theta}+3H\dot{\theta}=0,
\end{equation}
which implies that $\dot{\theta}=ba^{-3}$, $b$ is a constant of integration.
Using this result in Eq.(\ref{5}), we have
\begin{equation}\label{8}
H^{2}=\frac{1}{3}(\rho_{m}+\rho_{d})+\frac{1}{6}{b^{2}a^{-6}}.
\end{equation}

We consider the interacting scenario between DE and dark matter and
thus equation of continuity turns to the
following equations
\begin{eqnarray}
\dot{\rho}_{m}+3H\rho_{m}&=&Q, \label{9}
\\\label{10}\dot{\rho}_{d}+3H(\rho_{d}+p_{d})&=&-Q.
\end{eqnarray}
Here, $\rho_{d}$ is the energy density of the DE and $\rho_{m}$ is
the energy density of the pressureless matter and $Q$ is the
interaction term. Basically, $Q$ represents the rate of energy
exchange between DE and dark matter. If $Q>0$, it shows that energy
is being transferred from DE to the dark matter. For $Q<0$, the
energy is being transferred from dark matter to the DE. We consider
a specific form of interaction which is defined as $Q=3 H d^{2}
\rho_{m}$ and $d^{2}$ is interacting parameter which shows the
energy  transfers between CDM and DE. If we take $d=0$, then it
shows that each components, that is the non-relativistic matter and
DE, are self conserved. Using the value of $Q$ in Eq. (\ref{9}) we
have
\begin{equation}\label{11}
    \rho_{m}= \rho_{m 0}a^{-3(1-d^{2})},
\end{equation}
where, $\rho_{m0}$ is an integration constant. Hence, Eq.(\ref{10})
finally leads to the expression for pressure as follows
\begin{equation}\label{12}
p_{d}=-\left({d}^{2}{\rho_{m}+\rho_{d}}+\frac{\dot{\rho}_{d}}{3H}\right),
\end{equation}

The \textbf{\underline{EoS parameter}} is used to categorized the
decelerated and accelerated phases of the universe. This parameter
is defined as
\begin{equation}\label{13}
\omega=\frac{p}{\rho}.
\end{equation}
If we take $\omega=0$, it corresponds to non-relativistic matter and
the decelerated phase of the universe involve radiation era $0
<\omega<\frac{1}{3}$. For $\omega=-1, -1 <\omega<-\frac{1}{3} $ and
$\omega <-1$ correspond to the cosmological constant, quintessence
and phantom eras respectively. To analyze the dynamical properties
of the DE models, we use \underline{\textbf{$\omega-\omega'$ plane}}
\cite{27}. This plane describes the evolutionary universe with two
different cases freezing region and thawing region. In the freezing
region the values of EoS parameter and evolutionary parameter are
negative ($\omega<0$ and $\omega'<0$) while for the thawing region,
the value of EoS parameter is negative and evolutionary parameter is
positive ($\omega<0$ and $\omega'>0$). In order to check the
stability of the DE models, we need to evaluate the
\textbf{\underline{squared sound speed}} which is given by
\begin{equation}\label{14}
v_{s}^{2}=\frac{d p}{d \rho}=\frac{d p/dt}{d\rho/dt}.
\end{equation}
The sign of $v_{s}^{2}$ decides its stability of DE models, when
$v_{s}^{2}>0 $ the model is stable otherwise it is unstable.

\section{Tsallis Holographic Dark Energy}

The definition and derivation of standard HDE density is given by
$\rho_{d}=3c^{2}{m_{p}}^{2}/L^{2}$, where ${m_{p}}^{2}$ represents
reduced Plank mass and $L$ denotes the infrared cut-off. It depends
upon the entropy area relationship of black holes i.e $S\sim A\sim
L^{2}$ , where $A=4{\pi}L^{2}$ represents the area of the horizon.
Tsallis and Cirto \cite{32} studied that the horizon entropy of the
black hole can be modified as
\begin{equation}\label{15}
S_{\delta}=\gamma A^{\delta},
\end{equation}
where $\delta$ is the non-additivity parameter and $\gamma$ is an
unknown constant \cite{32}. Cohen at al. \cite{33}, proposed the
mutual relationship between IR (L) cut-off, system entropy (S) and
UV $(\Lambda)$ cut-off as
\begin{equation}\label{16}
L^{3}\Lambda^{3} \leq (S)^{\frac{3}{4}}.
\end{equation}
After combining  Eqs.(\ref{15}) and (\ref{16}), we get the following relation
\begin{equation}\label{17}
\Lambda^{4} \leq \gamma(4 \pi)^{\delta}L^{2\delta-4},
\end{equation}
where $\Lambda^{4}$ is vacuum energy density and $\rho_{d} \sim \Lambda^{4}$.
So, the Tsallis HDE density \cite{24} is given as:
\begin{equation}\label{18}
\rho_{d}=BL^{2\delta-4}.
\end{equation}
Here, $B$ is an unknown parameter and IR cutoff is taken as Hubble
radius which leads to  $L=\frac{1}{H}$, where $H$ is Hubble
parameter. The density of Tsallis HDE model along with its
derivative by using Eq.(\ref{18}) become
\begin{equation}
\rho_{d}=B{H^{4-2\delta}},\label{19} \quad \dot{\rho}_{d}= B (4-2\delta ) H^{3-2 \delta } \dot{H}.
\end{equation}
Where, $\dot{H}$ is the derivative of Hubble parameter w.r.t $t$.
The value of $\dot{H}$ is calculated in terms of $z$ using $a=\frac{1}{1+z}$ which is given as
\begin{eqnarray}
  \frac{dH}{dz} &=& \frac{\frac{1}{2}\left(\rho_{m 0}(1-d^2) (1+z)^{3(1-d^2)}+b^2(1+z)^6\right)}{\left(1-\frac{1}{3}B(4-2\delta )H^{3-2\delta }\right)H (1+z)}
\end{eqnarray}
Inserting these values in Eq.(\ref{12}) it yields
\begin{equation}\label{21}
p_{d}=\frac{1}{3} \left(-3 d^2 \rho_{m 0} a^{-3(1- d^2)}-B H^{2-2 \delta } \left(3 H^2+ (4-2\delta ) \dot{H}\right)\right).
\end{equation}
The EoS is obtained from Eq.(\ref{13})
\begin{equation}\label{22}
\omega_{d}=\frac{p_{d}}{\rho_{d}}=-1-\frac{d^2 \rho_{m 0} a^{-3(1- d^2)} H^{2 \delta-4 }}{B}+\frac{ (2\delta-4 ) \dot{H}}{3 H^2}.
\end{equation}
The plot of $\omega_{d}$ versus $z$ is shown in Figure \textbf{1}.
In this parameter and further results, the function $H(z)$ is being
utilized numerically. The other constant parameters are mentioned in
the Figure \textbf{1}. The trajectory of EoS parameter remains in
quintessence region at early, present and latter epoch.
\begin{figure}[htpb]
\centering \epsfig{file=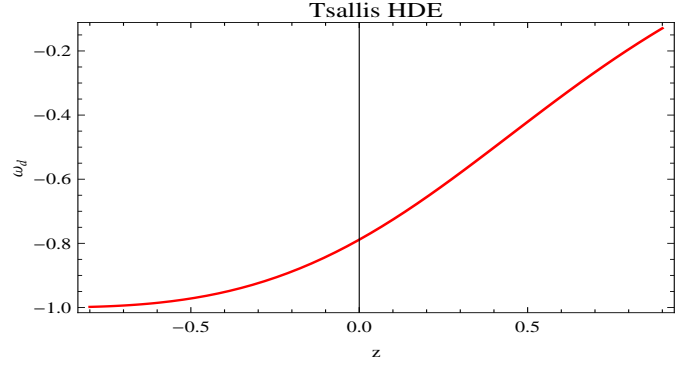, width=.65\linewidth,
height=1.9in} \caption {Plot of $\omega_{d}$ versus $z$ for THDE
model where $ \delta = 1.1, \rho_{m0}= 1, d^{2} = 0.001, B = -1.3, b
= 0.5.$}
\end{figure}

The square of the sound speed is given by
\begin{eqnarray}\nonumber
v_{s}^{2}&=& \frac{1}{6 B (\delta-2 ) a^4 H^3 \dot{H}}\left(9 d^2 \left(d^2-1\right)
\rho_{m0} a^{3 d^2} H^{2 \delta } \dot{a}-2 B (\delta-2 ) a^4 H\right.\\
\label{23} &\times&\left.\left(3 H^2 \dot{H}-2 (\delta-1 ) \dot{H}^2+H \ddot{H}\right)\right).
\end{eqnarray}
\begin{figure}[htpb]
\centering \epsfig{file=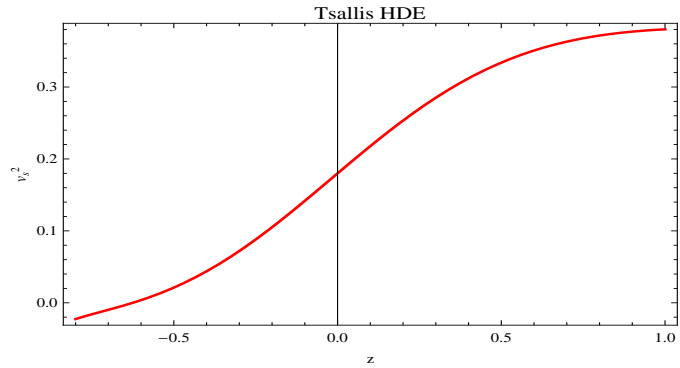, width=.65\linewidth,
height=1.9in} \caption {Plot of $v_{s}^{2}$ versus $z$ for THDE
model where $ \delta = 1.1, \rho_{m0}= 0.8, d^{2} = 0.001, B = -1.3,
b = 0.5.$}
\end{figure}
The plot of squared sound speed versus $z$ shown in Figure
\textbf{2} for different parametric values. This graph is used to
analyze the stability of this model. We can see that $v_{s}^{2}>0$,
for $-0.6 < z <1$ which corresponds to the stability of THDE model.
However, model shows instability for $z <-0.6$.

Taking the derivative of the EoS parameter with respect to  $\ln a$,
we get $\omega '_{d}$  as follows:
\begin{eqnarray}\nonumber
\omega '_{d} &=&  \frac{1}{3 B a^4 H^6}\left(-3 d^2 \rho_{m0} ^{3 d^2} H^{2 \delta }
\left(3 \left(d^2-1\right) H \dot{a}+ (2\delta-4 ) \dot{H}\right)+2 B (\delta-2 ) \right.\\
\label{24}&\times&\left. a^4 H^2\left(-2 \dot{H}^2+H \ddot{H}\right)\right)
\end{eqnarray}
\begin{figure}[htpb]
\centering \epsfig{file=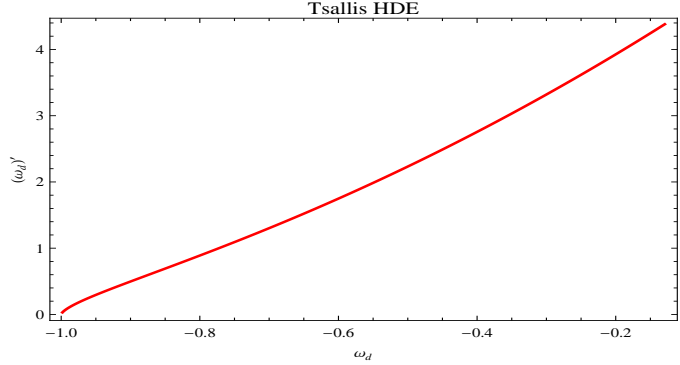, width=.65\linewidth,
height=1.9in} \caption {Plot of $\omega_{d}$ versus $\omega_{d}'$
for THDE model where $ \delta = 1.1, \rho_{m0}= 1, d^{2} = 0.001, B
= -1.3, b = 0.5.$}
\end{figure}
The graph of $\omega_{d}$ versus $\omega_{d}'$ is shown in Figure
\textbf{3}, for which $\omega_{d}'$ depicts positive behavior.
Hence, for $\omega_{d}<0$, the evolution parameter shows
$\omega_{d}'>0$, which represents the thawing region of evolving
universe.

\section{R\'{e}nyi Holographic Dark Energy Model}

We consider a system with $W$ states with probability of getting
$i^{\textmd{th}}$ state $P_{i}$ and satisfies the condition
$\Sigma{_{i=1}^{W}} P_{i}=1$. R\'{e}nyi and Tsallis entropies are
defined as
\begin{eqnarray}\label{24}
\mathcal{S}=\frac{1}{\delta} \ln \Sigma{_{i=1}^{W}}
P{_{i}^{1-\delta}},\quad
S_{T}=\frac{1}{\delta}\Sigma{_{i=1}^{W}}(P{_{i}^{1-\delta}}-P_{i}),
\end{eqnarray}
where $\delta\equiv 1-U$, where, $U$ is a real parameter. Now,
combining
 above equations we find  their mutual relation given as
\begin{equation}\label{26}
\mathcal{S}=\frac{1}{\delta} \ln (1+\delta S_{T}).
\end{equation}
This equation shows that $\mathcal{S}$ belongs to the class of most
general entropy functions of homogenous system. Recently, it has
been observed that Bekenstine entropy, $S=\frac{A}{4}$ is in fact
Tsallis entropy which gives the expression
\begin{equation}\label{27}
\textsl{S}=\frac{1}{\delta} \ln (1+\delta \frac{A}{4}),
\end{equation}\
which is the R\'{e}nyi entropy of the system. Now for the RHDE, we
focus on WMAP data for flat universe. Using the assumption
$\rho_{d}dv\propto Tds$, we can get RHDE density
\begin{equation}\label{28}
\rho_{d}=\frac{3 C^{2} H^{2}}{8\pi(1+\frac{\delta\pi}{H^{2}})}.
\end{equation}
Consider the term $8\pi=1$ substituting  in
Eq.(\ref{28}), we get the expression for density as
\begin{equation}\label{29}
\rho_{d}=\frac{3 C^{2} H^{2}}{1+\frac{\delta\pi}{H^{2}}}.
\end{equation}
Now, $\frac{dH}{dz}$  is given by
\begin{eqnarray}
\frac{dH}{dz}  &=&
\frac{\frac{1}{2}\left(\rho_{m0}(1-d^2)(1+z)^{3(1-d^2)}+b^2(1+z)^6\right)}{\left(1-\frac{2c^2H^2\left(z^2+\delta
\pi \right)-c^2H^4}{\left(H^2+\delta  \pi \right)^2}\right)H(1+z)}
\end{eqnarray}
The pressure for this case is obtained as
\begin{equation}\label{30}
p_{d}=-d^2 \rho_{m0} a^{-3(1- d^2)}+\frac{c^2 H^2 \left(-3 H^2 \left(\pi  \delta
+H^2\right)-2 \left(2 \pi  \delta +H^2\right) \dot{H}\right)}{\left(\pi  \delta +H^2\right)^2}.
\end{equation}
The expressions for EoS parameter $\omega_{d}$ can be evaluated from
Eq.(\ref{12}) as follows
\begin{equation}\label{31}
\omega_{d}=\left(\pi  \delta +H^2\right) \left(\frac{-d^2 \rho_{m0} a^{-3(1- d^2)}}{3 c^2 H^4}
-\frac{\left(3 H^2 \left(\pi  \delta +H^2\right)+2 \left(2 \pi  \delta +H^2\right) \dot{H}\right)}{3H^2\left(\pi  \delta +H^2\right)^2}\right).
\end{equation}
\begin{figure}[htpb]
\centering \epsfig{file=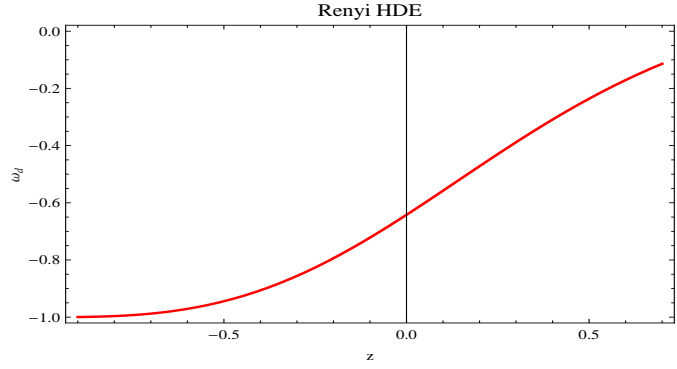, width=.65\linewidth,
height=1.9in} \caption {Plot of $\omega_{d}$ versus $z$ for RHDE
model where $ \delta = 1.1, \rho_{m0}= 0.8, d^{2} = 0.001, c = 0.1,
b = 0.05.$}
\end{figure}

Figure \textbf{4} shows the plot of $\omega_{d}$ versus $z$. The
trajectory of EoS parameter evolutes the universe from quintessence
region towards the $\Lambda$CDM limit. The squared sound speed of
this RHDE model is given by using Eq.(\ref{13}) as
\begin{eqnarray}
\nonumber{v_s}^{2}&=&\frac{3 H\left(1-d^2\right) d^2\rho_{m0}  a^{-3(1- d^2)}\left(\pi  \delta +H^2\right)^2  }{6 c^2 H^3 \left(2 \pi  \delta +H^2\right) \dot{H}} -\frac{1}{3H^2 \left(2 \pi  \delta +H^2\right)\left(\pi  \delta +H^2\right)}\\
\nonumber &\times&\left\{ \dot{H} \left(6 \pi ^2 \delta ^2 H^2+9 \pi  \delta  H^4+3 H^6+4 \pi ^2 \delta ^2 \dot{H}\right)+ H\ddot{H} \left(\pi  \delta +H^2\right)\right. \\
\label{32} &\times&\left.\left(2 \pi  \delta +H^2\right) \right\}.
\end{eqnarray}
\begin{figure}[htpb]
\centering \epsfig{file=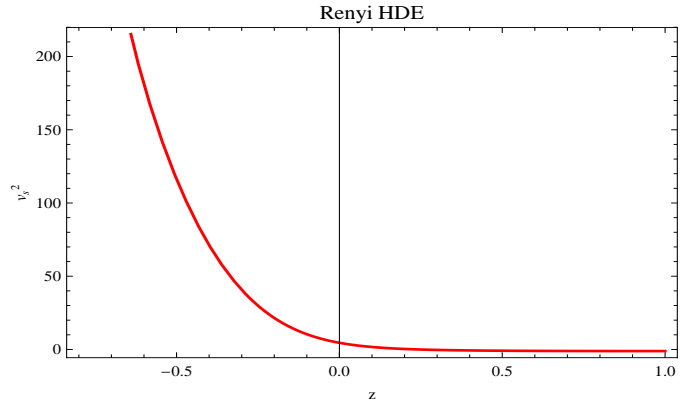, width=.65\linewidth,
height=2.1in} \caption {Plot of $v_{s}^{2}$ versus $z$, for RHDE
model where $ \delta = 1.1, \rho_{m0}= 0.8, d^{2} = 0.001, c=0.1, b
= 1.5.$}
\end{figure}

The graph of squared speed of sound is shown in Figure \textbf{5}
versus $z$. In this case, we have $v_{s}^{2}>0$ for all range of $z$
which shows the stability of RHDE model at the early, present and
latter epoch of the universe.

The expression for $\omega'_{d}$ is evaluated as:
\begin{eqnarray}\nonumber
\omega'_{d} &=&  \frac{1}{3 c^2 a^4 H^6 \left(\pi  \delta +H^2\right)^2}\left\{-d^2
\rho_{m0} a^{3 d^2} \left(\pi  \delta +H^2\right)^2 \left(3H\dot{a} \left(-1+d^2\right)\right.\right.\\
\nonumber &\times& \left.\left. \left(\pi  \delta +H^2\right) - 2 a\dot{H} \left(2 \pi  \delta +H^2\right)
\right)+2 c^2 a^4 H^2  \left(4 \pi ^2 \delta ^2+8 \pi  \delta  H^2+2H^4\right) \dot{H}^2\right.\\
\label{31} &-&\left.2H \left(\pi  \delta +H^2\right) \left(2 \pi  \delta +H^2\right) \ddot{H}\right\}.
\end{eqnarray}
\begin{figure}[htpb]
\centering \epsfig{file=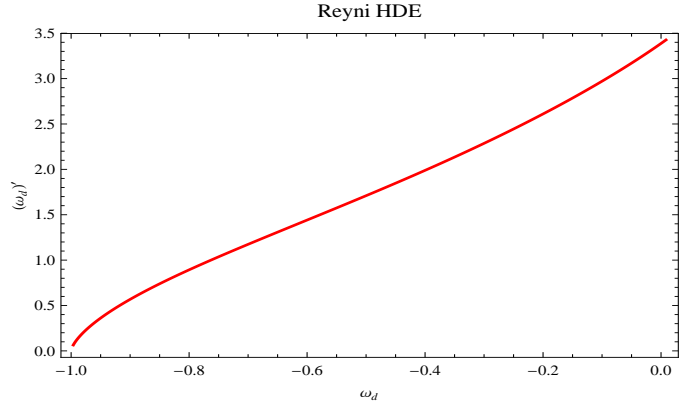, width=.65\linewidth,
height=2.1in} \caption {Plot of $\omega_{d}$ versus $\omega_{d}'$
for RHDE model where $ \delta =1.1, \rho_{m0}= 0.8, d^{2}=0.001,
c=0.1,b=0.05.$}
\end{figure}
In Figure (6), we plot the EoS parameter with its evolution
parameter to discuss $\omega_{d}-\omega_{d}'$ plane for RHDE model.
The graph shows that for $\omega_{d}<0$, the evolutionary parameter
remains positive at the early, present and latter epoch. This type
of behavior depicts the thawing region of the evolving universe.

\section{Sharma-Mitall Holographic Dark Energy Model}

From the R\'{e}nyi entropy, we have the generalized entropy content
of the system. Using Eq.(\ref{26}), Sharma-Mittal introduced a two
parametric entropy which is defined as
\begin{equation}\label{33}
S_{SM}=\frac{1}{1-r}\left((\Sigma{_{i=1}^{W}} P{_{i}^{1-\delta}})^{1-r/\delta}-1\right),
\end{equation}
where $r$ is a new free parameter. We can observe that R\'{e}nyi and
Tsallis entropies can be recovered at the  proper limits, using
Eq.(\ref{24}) in Eq.(\ref{33}), we have
\begin{equation}\label{34}
S_{SM}=\frac{1}{R}((1+\delta S_{T})^{R/\delta}-1),
\end{equation}
here, $R \equiv 1-r $. Using the argument that Bekenstine
entropy is the proper candidate for Tsallis entropy by using $S=A/4$
where $A$ is horizon entropy, we get the following expression
\begin{equation}\label{35}
S_{SM}=\frac{1}{R}((1+\delta \frac{A}{4})^{R/\delta}-1),
\end{equation}
The relation of UV $(\Lambda)$ cut off, IR (L) cut off and and system horizon (S) is given as
\begin{equation}\label{36}
\Lambda^{4} \propto \frac{S}{L^{4}}
\end{equation}

Now, taking $L \equiv \frac{1}{H}=\sqrt{A/4\pi}$, then the the
energy density of DE given by Sharma-Mitall \cite{24} is considered
as;
\begin{equation}\label{37}
\rho_{d}=\frac{3c^{2}H^{4}}{8\pi R}\left[(1+\frac{\delta \pi}{H^{2}})^{R/\delta}-1\right],
\end{equation}
here, $c^{2}$  is an unknown free parameter. Using
$8\pi=1$ in above equation, we get the following expression for
energy density
\begin{equation}\label{38}
\rho_{d}=\frac{3c^{2}H^{4}}{ R}\left[\left(1+\frac{\delta \pi}{H^{2}}\right)^{R/\delta}-1\right].
\end{equation}
The differential equation of $H$ is given by
\begin{eqnarray}
\frac{dH}{dz} &=&
\frac{\frac{1}{2}\left(\rho_{m0}(1-d^2)(1+z)^{3(1-d^2)}+b^2(1+z)^6\right)}{1+c^2
\pi \left(1+\frac{\delta  \pi }{H^2}\right)^{\frac{R}{\delta
}-1}-\frac{2 c^2 H^2}{R}\left(\left(1+\frac{\delta  \pi
}{H^2}\right)^{\frac{R}{\delta }}-1\right)H(1+z)}
\end{eqnarray}
The pressure can be evaluated by energy conservation Eq.(\ref{11})
as follows
\begin{eqnarray}\label{39}
\nonumber p_{d}&=&-d^2 \rho_{m0} a^{-3(1- d^2)}-c^2 \left(\frac{3 \left(\left(1+\frac{\pi  \delta }{H^2}\right)^{R/\delta }-1\right) H^4}{R}-2 \pi \dot{H}  \left(1+\frac{\pi  \delta }{H^2}\right)^{{R/\delta }-1}\right.\\
\nonumber &+& \left.\frac{4 \left(\left(1+\frac{\pi  \delta }{H^2}\right)^{R/\delta }-1\right) H^2 \dot{H}}{R}\right).
\end{eqnarray}
The EoS parameter for this model is given by
\begin{eqnarray}\label{40}
\nonumber \omega_{d}&=&2 c^2  \left(\pi  \left(1+\frac{\pi  \delta }{H^2}\right)^{{R/\delta}-1}-\frac{2H^2\dot{H}}{R}
 \left(\left(1+\frac{\pi  \delta }{H^2}-1\right)^{R/\delta }\right) \right)\\
\nonumber&-&\frac{d^2 R \rho_{m0} a^{-3(1- d^2)}}{3c^2 H^4\left(\left(1+ \frac{\delta  \pi }{H^2}\right)^{{R/\delta }}-1\right)} -1.
\end{eqnarray}
\begin{figure}[htpb]
\centering \epsfig{file=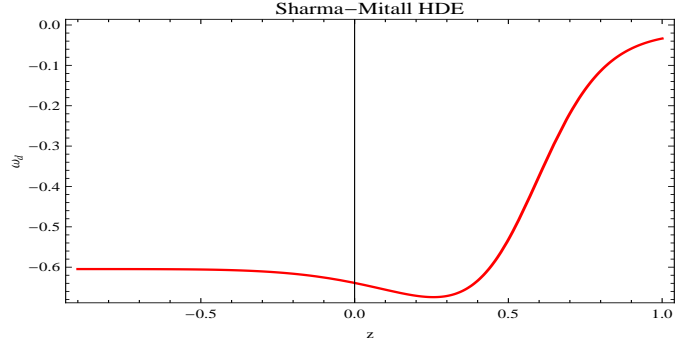, width=.65\linewidth,
height=1.8in} \caption {Plot of $\omega_{d}$ versus $z$ for SMHDE
where $ \delta =1.1, \rho_{m0}= 0.01, d^{2}=0.001, c=0.01, b=0.4,
R=7.$}
\end{figure}
The plot of $\omega_{d}$ versus $z$ is shown in Figure \textbf{7}.
The EoS parameter represents the quintessence nature of the
universe. The square of the sound speed is evaluated as
\begin{eqnarray}\nonumber
\nonumber v_{s}^{2}&=&\frac{1}{6 c^2 H\dot{H} \left(-\pi
\left(1+\frac{\pi  \delta }{H^2}\right)^{R/\delta-1 }+\frac{2H^2}{R} \left(\left(1+\frac{\pi  \delta }{H^2}\right)^{R/\delta}-1\right) \right) }\\
\nonumber&\times& \left\{-3 d^2 H\left(-1+d^2\right) \rho_{m0} a^{-3(1- d^2)}+\frac{2 c^2 H}{R} \left(6 H^2 \dot{H}+4 \dot{H}^2+2 H \ddot{H}\right.\right.\\
\nonumber&-&\left.\left.\frac{1}{\left(\pi  \delta +H^2\right)^2}\left(1+\frac{\pi  \delta }{H^2}\right)^{R/\delta }  \left(3 H^2\dot{H} \left(\pi  \delta +H^2\right) \left(-\pi  R+2 \pi  \delta +2 H^2\right)\right.\right.\right.\\
\nonumber&+&\left.\left.\left.2\dot{H} \left(\pi ^2 (R-2 \delta ) (R-\delta )-2\dot{H}^2 \pi  (R-2 \delta ) H^2+2 H^4\right) \right)+H\dot{H} \left(\pi  \delta +H^2\right)\right.\right.\\
\label{41}&\times&\left.\left.\left(-\pi  R
+2 \pi  \delta +2 H^2\right) \ddot{H}\right)\right\}.
\end{eqnarray}

\begin{figure}[htpb]
\centering \epsfig{file=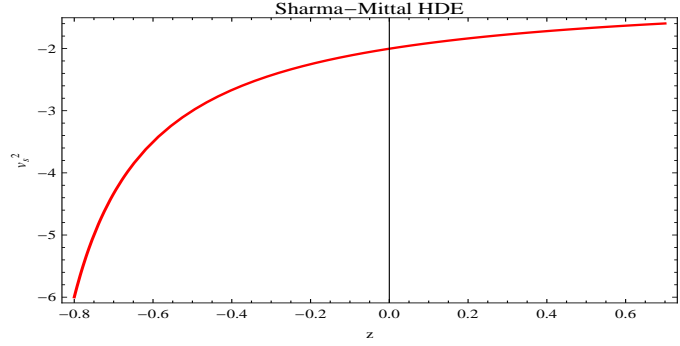, width=.65\linewidth,
height=1.8in} \caption {Plot of $v_{s}^{2}$ versus $z$ for SMHDE
where $ \delta =1.1, \rho_{m0}= 0.8, d^{2}=0.001, c=0.8, b=0.05,
R=7.$}
\end{figure}

In Figure 8, we draw $v_{s}^{2}$ versus $z$ which shows the
un-stable behavior of the SMHDE model as $v_{s}^{2}<0$ at early,
present and latter epoch.

\begin{eqnarray}
\omega'_{d} &=& -\frac{1}{3\left(\left(1+\frac{\pi  \delta }{H^2}\right)^{R/\delta }-1\right)^2 H^6}\left(\frac{1}{\left(\pi\delta +H^2\right)^2}2 H^2 \left(2 \left(-2 \left(\pi  \delta +H^2\right)^2\right.\right.\right.\\
\nonumber&+&\left.\left.\left.\left(1+\frac{\pi  \delta }{H^2}\right)^{2 R/\delta } \left(\pi ^2 (R-2 \delta ) \delta +2 \pi  (R-2 \delta ) H^2-2 H^4\right)+\left(1+\frac{\pi  \delta }{H^2}\right)^{R/\delta }\right.\right.\right.\\
\nonumber&\times&\left.\left.\left. \left(-\pi ^2 \left(R^2+R \delta -4 \delta ^2\right)-2 \pi(R-4 \delta ) H^2+4 H^4\right)\right)\dot{H}^2+\left(\pi  \delta +H^2\right)\right.\right.\\
\nonumber &\times&\left.\left.\left(\left(1+\frac{\pi  \delta }{H^2}\right)^{R/\delta}-1\right)H \left(-2 \left(\pi  \delta +H^2\right)+\left(1+\frac{\pi  \delta}{H^2}\right)^{R/\delta}\right.\right.\right.\\
\nonumber&\times&\left.\left.\left.\left(-\pi R+2 \pi\delta+2
H^2\right)\right)\ddot{H}\right)
+\frac{3 d^2 \left(-1+d^2\right)}{c^2} \rho_{m0} R a^{-3(1- d^2)}H^2\right. \\
\nonumber&\times&\left.\left(\left(1+\frac{\pi  \delta }{H^2}\right)^{R/\delta }-1\right)+\frac{2 d^2 \rho_{m0} R a^{-3(1- d^2)}}{c^2 \left(\pi  \delta +H^2\right)}\left( \left(\pi  (R-2 \delta )-2 H^2\right)\right.\right.\\
\nonumber&\times&\left.\left. \left(1+\frac{\pi  \delta }{H^2}\right)^{R/\delta }+2 \left(\pi  \delta +H^2\right)\right) \dot{H}\right)
\end{eqnarray}
\begin{figure}[htpb]
\centering \epsfig{file=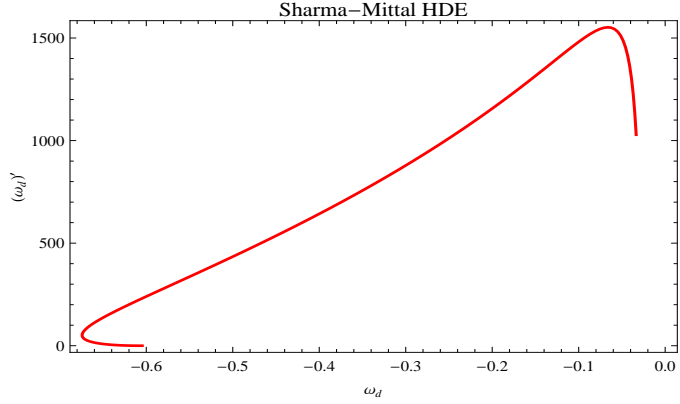, width=.65\linewidth,
height=2.1in} \caption {Plot of $\omega_{d}$ versus $\omega_{d}'$
for different values of $\delta$ for SMHDE where $ \delta =1.1,
\rho_{m0}= 0.01, d^{2}=0.001, c=0.01, b=0.4, R=7.$}
\end{figure}

Figure \textbf{9} shows the plot of $\omega_{d}$-$\omega_{d}'$ plane
to classify the dynamical region for the given model. We can see
that, $\omega_{d}'>0$ for $\omega_{d}<0$, which indicates the
thawing region of the universe.

\section{Conclusion}

In this paper, we have discussed the THDE, RHDE and SMHDE models
in the frame work of Chern-Simons modified theory of gravity. We
have taken the flat FRW universe and linear
interaction term is chosen for the interacting scenario between DE
and dark matter. We have evaluated the different cosmological
parameters (equation of state parameter and squared sound speed),
$\omega_{d}- \omega_{d}'$ cosmological plane. The trajectories of
all these models have been plotted with different constant
parametric values.

We have summarized our results in the following table.\\
\textbf{Table 1:} Summary of the cosmological parameters and plane.
\begin{table}[bht]
\centering
\begin{small}
\begin{tabular}{|c|c|c|c|c|c|}
\hline DE models & $\omega_d$ & $v^2_s$ & $\omega_d-\omega_d'$\\
\hline THDE & quintessence-to-vacuum & partially stability  & thawing region\\
\hline RHDE & quintessence-to-vacuum & stability  & thawing region\\
\hline SMHDE & quintessence & un-stable & thawing region\\
\hline
\end{tabular}
\end{small}
\end{table}\\\\

Jawad et al. \cite{38j} have explored various cosmological
parameters (equation of state, squared speed of sound,
Om-diagnostic) and cosmological planes in the framework of dynamical
Chern-Simons modified gravity with the new holographic dark energy
model. They observed that the equation of state parameter gives
consistent ranges by using different observational schemes. They
also found that the squared speed of sound shows a stable solution.
They suggested that the results of cosmological parameters show
consistency with recent observational data. Jawad et al. \cite{39j}
have also considered the power law and the entropy corrected HDE
models with Hubble horizon in the dynamical Chern–Simons modified
gravity. They have also explored various cosmological parameters and
planes and found consistent results with observational data. Nadeem
et al. \cite{40j} have also investigated the interacting modified
QCD ghost DE and generalized ghost pilgrim DE with cold dark matter
in the framework of dynamical Chern-Simons modified gravity. It is
found that the results of cosmological parameters as well as planes
explain the accelerated expansion of the Universe and are compatible
with observational data.

However, the present work is different from the above mentioned
works in which we have taken recently proposed DE models along with
non-linear interaction term and found interesting and compatible
results regarding current accelerated expansion of the universe.

\section*{Data Availability}

The data used to support the findings of this study are available
from the corresponding author upon request.

\section*{Acknowledgments}
The authors declare that there is no conflict of interest regarding
the publication of this paper. Also, the work of H. Moradpour has
been supported financially by Research Institute for Astronomy \&
Astrophysics of Maragha (RIAAM) under research project No.
$1/5237-8$. The mentioned received funding in the "Acknowledgment"
section did not lead to any conflict of interests regarding the
publication of this manuscript.


\vspace{.25cm}
\end{document}